\documentclass[12pt]{article}
\usepackage{amssymb,amsmath,epsfig}

\begin{document}

\title{\bf Dynamics of Charged Viscous Dissipative Cylindrical Collapse With Full Causal Approach}
\author{S.M. Shah and G. Abbas \thanks{abbasg91@yahoo.com}
\\Department of Mathematics, The Islamia University of Bahawalpur,\\Bahawalpur, Pakistan.}
\date{}
\maketitle
\begin{abstract}
The aim of this paper is to investigate the dynamical aspects of charged viscous cylindrical source by using Misner approach. To this end, we have considered the more general charged dissipative fluid enclosed by the cylindrical symmetric spacetime. The dissipative nature of the source is due to the presence of dissipative variables in the stress-energy tensor. The dynamical equations resulting from such charged cylindrical dissipative source have been coupled with the causal transport equations for heat flux, shear and bulk viscosity, in the context of Israel-Steward theory. In this case, we have the considered the Israel-Steward transportation equations without excluding the thermodynamics viscous/heat coupling coefficients. The results are compared with the previous works in which such coefficients were excluded and viscosity variables do not satisfy the casual transportation equations.
\end{abstract}
{\bf Keywords:} Gravitational
Collapse; Israel-Steward theory, Full Casual Approach. \\
 {\bf PACS:} 04.20.Cv; 04.20.Dw
\section{Introduction}
An important and renowned aspect of general relativity can be observed in gravitational collapse of sufficiently heavy stars, having mass greater than the mass of sun. This phenomena occurs due to the positive difference of gravity and pressure, in the internal nuclear forces of the massive stars. Openheimer and Snyder   \cite{1} are innovators of this field of research. They take an initiative step in the field of gravitational collapse and discussed a detailed note on the subject of "On continued gravitational contraction". On the other hand an intellectual and substantial investigation was contributed by Misner and Sharp \cite{2} in 1964. They considered perfect fluid inside stars and formulated dynamical equations for adiabatic relativistic collapse. Vaidya \cite{3} find the exact model of collapse for the radiating star and some researchers \cite{4}-\cite{16} investigated the gravitational collapse in different situations. According to Rosssland \cite{17} ions are conversion of atoms having large strength and explained that in between free particles, the law of cental force can be detected, he further addressed that electrical forces have great effect for a star which has the mass $1.5 M_{\bigodot}$ (where $M_{\bigodot}$ is mass of the sun) and its molecular cloud has weight $2.8$ units.

Mitra \cite{18} observed that formation and evolution of stars is highly dissipative process, which can be divided into two phases. Phase one is free streaming approximation while second phase is streaming approximation. Tewari \cite{19}-\cite{21} find the solutions of Einstein field equations for different models. A lot of work for diffusion approximation with electromegnetic field, anisotropy, inhomogeneity and viscosity has been discovered by many well known relativists \cite{22}-\cite{25}. In 1987, it is investigated from supernova that regime of radiation is far from streaming out limit than the deffusion approximation \cite{26}. Arnett and Kazanas \cite{27}-\cite{28} also detected that the amount of vigor of radiation is directly related to the temperature gradient. Generally, this detection is strongly reasonable because the mean free path of particles being efficient causes for energy transfer is smaller than the typical objects. Hence for an important progression star as the sun, the mean free path of photons at the center is of order 2 cm. Also the mean free path of trapped neutrinos in compact core of densities about $10^{12} g.cm^{-3}$ becomes smaller than the size of the stellar core. Eckart \cite{29} and Landau \cite{30} discussed the transport equation for shear viscosity and theory of relativistic ir-reversible thermodynamics. As effects of viscosity plays a vital role in development of neutron stars, therefore the coefficient of shear viscosity may attains its value up to $10^{20} gcm^{-1}s^{-1}$ \cite{31}. While the coefficient of bulk viscosity have its maximum value is $10^{30}gcm^{-1}s^{-1}$ \cite{32} due to Urca reaction  in neutron stars and white dwarfs.

Alford and Blaschve \cite{33},\cite{34} assumed that the two color super conducting quark matter may acquire same or large values  for crossbreed stars \cite{35}. Recently, Sharif and his collaborators \cite{36}-\cite{38} attempted to discuss the dynamics of collapsing cylindrically source. They considered cylindrical spacetimes which resembling to spherical spacetime. Also, Herrera et al. \cite{39} used the full casual approach to investigate the dynamics of spherical dissipative fluid with the radial heat flux. This work is very interesting, informative and have a great impact on the study of many realistic astrophysical phenomena. Our present work is distinguish in the sense that we have taken the geometry of stars. In addition to this, we have introduced electromagnetic field in energy momentum tensor and discuss its effects. This work opens a new direction and have different attractive and meaning full results in the field of dissipative gravitational collapse.

The purpose of the present study is to discuss the dynamics of charged viscous cylindrical gravitational collapse in the framework of Misner formulism. The dissipative nature of the source is prescribed by dissipative variables. The dynamical equations resulting from such charged cylindrical dissipative source are then coupled with causal transport equations for heat flux, shear and bulk viscosity, in the context of Israel-Steward theory by including the thermodynamics viscous/heat coupling coefficients. The inclusion of these coefficients many regarded as much important in the non-uniform stellar models \cite{Maa} and viscosity variables satisfy the casual transportation equations. The plan of the paper is as follow: in the next section, we have presented the dynamical equations for charged viscous cylindrical source. In section 3, the causal transport equations are coupled with dynamical. Finally, the results of the paper have been summarized in the last section.

\section{Viscous Dissipative Fluid Enclosed by Cylindrical Stellar Object }
This section  deals with geometry of stellar object, charged dissipative source of matter and the equations of motion. The interior metric in cylindrical coordinates is given by
\begin{eqnarray}\label{1m}
ds^{2}_{-}=-A^2dt^{2}+B^2dr^{2}+R^{2}d\theta^{2}+dz^{2},
\end{eqnarray}
where the constraints on coordinates are following

$-\infty\leq t\leq\infty$, $-\infty\leq z\leq\infty$, $0\leq r$, $0\leq \theta\leq 2\pi$ and $A=A(r,t)$, $B=B(r,t)$ and $R=R(r,t)$.

The energy momentum tensor for charged viscous dissipative fluid is
\begin{eqnarray}\nonumber
T_{\alpha\beta}&=&(\mu+P+\Pi)V_{\alpha}V_{\beta}+(P+\Pi)g_{\alpha\beta}+q_{\alpha}V_{\beta}+q_{\beta}V_{\alpha}+\epsilon l_{\alpha}l_{\beta}+\pi_{\alpha\beta}\\\label{2}
&+&\frac{1}{4\pi}\left(F^{\gamma}_{\alpha}F_{\beta\gamma}-
\frac{1}{4}F^{\gamma\delta}F_{\gamma\delta}g_{\alpha\beta}\right),
\end{eqnarray}
where $\mu$ , $p$, $V_{\alpha}$, $\pi_{\alpha\beta}$, $\Pi$, $\epsilon$ and $q_{\alpha}$ are energy density, isotropic pressure, four velocity, shear viscosity, bulk viscosity, radiation density and radial heat flux, respectively. Also,
$F_{\alpha\beta}=-\phi_{\alpha,\beta}+\phi_{\beta,\alpha}$ is the
electromagnetic tensor tensor with potential $\phi_\alpha$. Moreover,
$l^{\alpha}$ is a radial null four vector. The above quantities satisfy the following relations
\begin{equation*}
l^{\alpha}V_{\alpha}=-1,\quad V^{\alpha}q_{\alpha}=0,\quad l^{\alpha}l_{\alpha}=0,\quad
V^{\alpha}V_{\alpha}=-1,\quad
V^{\alpha}q_{\alpha}=0, \quad \pi_{\alpha\beta}V^{\alpha}=0.
\end{equation*}
Further,
\begin{eqnarray}\nonumber
V^{\alpha}&=&A^{-1}\delta^{\alpha}_{0},\quad
q^{\alpha}=qB^{-1}\delta^{\alpha}_{1},\quad
V_{\alpha}=-X\delta^{0}_{\alpha},\quad
l^{\alpha}=A^{-1}{\delta}^{\alpha}_{0}+B^{-1}{\delta}^{\alpha}_{1} \quad and \\\label{4}
 \pi_{\alpha\beta}&=&\Omega(\chi_{\alpha}\chi_{\beta}-\frac{1}{3}h_{\alpha\beta}).
\end{eqnarray}
The Maxwell field equations are
\begin{equation}\quad\quad\quad\label{4}
F^{\alpha\beta}_{~~;\beta}={4\pi}J^{\alpha},\quad
F_{[\alpha\beta;\gamma]}=0,
\end{equation}
where $J_\alpha$ is the four-current. We assume that the following form of electromagnetic potential and current
\begin{equation*}
\phi_{\alpha}={\phi}{\delta^{0}_{\alpha}},\quad
J^{\beta}={\zeta}V^{\beta}.
\end{equation*}
Here $\zeta(r,t)$ and $\phi(r,t)$  are charge density and scalar potential, respectively.
\newline

The non-zero components of expansion scalar is
\begin{eqnarray}\label{5}
 \Theta&=&\frac{1}{A}\left(\frac{2\dot{B}}{B}+\frac{\dot{R}}{R}\right),
 \end{eqnarray}

where $\partial_t$= $^.$ and $\partial_r$= $'$ The field equations for the given source are
\begin{equation}\label{1f}
8\pi\left(\mu+\epsilon+\frac{\pi}{2}E^{2}\right)A^{2}=\frac{{\dot{B}}\dot{R}}{BR}+\left(\frac{A}{B}\right)^{2}\left(\frac{A'R'}{AR}-\frac{R''}{R}\right),
\end{equation}
\begin{equation}\label{2f}
8\pi(q+\epsilon)AB=\frac{{\dot{R'}}}{R}-\frac{\dot{B}R'}{BR}-\frac{\dot{R}A'}{AR},
\end{equation}
\begin{equation}\label{9f}
8\pi\left(p+\Pi+\epsilon+\frac{2}{3}\Omega+\frac{\pi}{2}E^{2}\right)B^{2}=\frac{A'R'}{AR}+\left(\frac{B}{A}\right)^{2}\left(-\frac{\ddot{R}}{R}+
\frac{\dot{A}\dot{R}}{AR}\right),
\end{equation}
\begin{equation}\label{4f}
8\pi\left(p+\Pi-\frac{\Omega}{3}-\frac{\pi}{2}E^{2}\right)R^{2}=\left(\frac{1}{AB}\right)\left(\frac{\dot{A}\dot{B}}{A^{2}}-\frac{A'B'}{B^{2}}
-\frac{\ddot{B}}{A}+\frac{A''}{B}\right),
\end{equation}
where $E=\frac{\hat{Q}(r)}{2\pi R}$ and $\hat{Q}(r)=4\pi \int^r_0 \zeta BRdr $.

The gravitational energy per specific length (also, known as C-energy) for cylindrical
symmetric spacetime is given as follows \cite{40,40a}

\begin{equation*}
\hat{E}(r,t)=\frac{\Big(1-l^{-2}\nabla^{\alpha}\tilde{r}~\nabla_{\alpha} \tilde{r}\Big)}{8}.
\end{equation*}

For cylindrically symmetric model with killing vectors, the
circumference radius $\rho$ and specific length $l$ and areal radius $\tilde{r}$ are defined as
follows \cite{40,40a}

\begin{equation*}
\rho^2={\xi}_{(1)\alpha}{{\xi}}^{\alpha}_{(1)},\quad\quad l^2={\xi}_{(2)\alpha}{{\xi}}^{\alpha}_{(2)},\quad\quad \tilde{r}=l\rho
\end{equation*}

The C-energy in the total interior region with the electromagnetic field \cite{41} is given by
\begin{equation}\label{9m}
m\left(r,t\right)=l\hat{E}(r,t)=\frac{l}{8} \left[1+\left(\frac{\dot{R}}{A}\right)^{2}-\left(\frac{R'}{B}\right)^{2}\right]+\frac{l^2\hat{Q}^{2}}{2R}.
\end{equation}

The collapsing fluid reside inside the non-static metric (1), therefore it must be matched to a
suitable exterior. If heat leaves the fluid across boundary surface then exterior region of the collapsing star will not be vacuum, but
the outgoing Vaidya like spacetime which models the radiation and has metric \cite{42a}
\begin{equation}\label{2m}
ds^{2}_{+}=-\left(-\frac{2M(\nu)}{R}+\frac{\tilde{q}^{2}(\nu)}{R^{2}}\right)d\nu^{2}-2d\nu dR + R^{2}(d\theta^{2}+\lambda^{2}d\phi^2),
\end{equation}
where $M(\nu)$ and $\tilde{q}(\nu)$ are mass and charge respectively, these both are measured in unit length (as we have used the relativistic units in our calculations). Also, $\lambda$ is arbitrary constant having the unit of length. It has been introduced to balance the units in the metric.
Using the continuity of extrinsic curvature of the spacetimes given in Eq.(\ref{1m}) and Eq.(\ref{2m}), we get
 \begin{equation}
 M(\nu)=\frac{R}{2}\left[\left(\frac{\dot{R}}{A}\right)-\left(\frac{\acute{R}}{B}\right)\right]+\frac{\tilde{q}}{2R}
 \end{equation}
\begin{equation}
\acute{E}-M=^{\Sigma}\frac{1}{8}, \quad\quad  p+\Pi+ \frac{2}{3}\Omega=^{\Sigma}q.
 \end{equation}
 where $s=^{\Sigma}\tilde{q}$ has been used. These are the conditions for the smooth matching of two regions. Here,  $p+\Pi+ \frac{2}{3}\Omega=^{\Sigma}q$, implies that effective pressure on the boundary of the cylinder is non-zero and it is equal to radial heat flux which provide the possibility of gravitational radiation produced by the collapsing fluid.

\section{Dynamical Equations} By using  Misner and Sharp\cite{2},\cite{42} concept, we may be able to observe the dynamical behavior of field equations. So, we define $D_{t}$ as the proper time derivative of the following form
\begin{equation}\label{9d}
D_{t}=\frac{1}{A}\frac{\partial}{\partial t}.
\end{equation}
The velocity $U$ is
 \quad\quad\quad\quad\quad\quad\quad\quad\quad U=$D_{t}B$$<0$ \quad\quad\quad\quad\quad\quad(Collapse).
\newline Also from Eq.(\ref{9m}),we have
\begin{equation}\label{9}
\frac{R'}{B}=\left(1+U^{2}-\frac{8m}{l}+\frac{4\hat{Q}^{2}l}{R}\right)^{\frac{1}{2}}=\hat{E}.
\end{equation}
The proper time derivative of mass function is
\begin{equation}\label{9}
D_{t}m=l\left(\frac{\dot{R}\ddot{R}}{4A^{3}}-\frac{\dot{R^{2}}\dot{A}}{BA^{4}}-\frac{R'\dot{R'}}{4B^{2}A}+\frac{R'^{2}\dot{B}}{4AB^{3}}\right)
-\frac{\dot{R}\hat{Q}^{2}l^{2}}{2AR^{2}}.
\end{equation}
Using Eq.(6), Eq.(7) and the value of $E=$$\frac{\hat{Q}}{2\pi R}$ then above equation takes the form
\begin{equation}\label{9}
D_{t}m=-2\pi l\left(\hat{E}(q+\epsilon)B+U(p+\Pi+\epsilon+\frac{2}{3}\Omega)\right)R.
\end{equation}
Above relation leads us about the variation rate of energy in the cylinder of radius $R$ and right hand side of this equation describes increment in energy in the interior region of radius $R$. The term $(p+\Pi+\frac{2}{3}\Omega)$ is effective radial pressure while $\epsilon$ denotes the pressure of radiation. The standard thermodynamical relation is $\pi_{\alpha \beta}$ in static phase and first term in R.H.S of above equation demonstrates energy of source, which is leaving the cylindrical surface.
\newline The dynamics of collapsing system can be observe through the proper radial derivative $D_{R}$ , which is defined as follows

\begin{equation}\label{9}
D_{R}=\frac{1}{R'}\frac{\partial}{\partial r}.
\end{equation}
Substituting Eq.(17) in Eq.(09), we have

\begin{eqnarray}\label{9}
D_{R}m&=&\frac{l}{R'}\left[\frac{\dot{R}\dot{R'}}{4A^{2}}-\frac{\dot{R}^{2}A'}{4A^{3}}-\frac{R'R''}{4B^{2}}
+\frac{B'R'^{2}}{4B^{3}}+\frac{l\hat{Q}\hat{Q}'}{R}-\frac{l\hat{Q}^{2}R'}{2R^{2}}\right].
\end{eqnarray}

Substituting Eq.(5) and Eq.(6) in Eq.(17), we have
\begin{equation}\label{9aaaa}
D_{R}m=2\pi Rl\left(4(\mu+\epsilon)+\frac{U}{\hat{E}}(q+\epsilon)B\right)+\frac{l^{2}\hat{Q}\hat{Q}'}{RR'}-\frac{l^{2}\hat{Q}^{2}}{R^{2}}.
\end{equation}
The above equation on integration yields
\begin{equation}\label{9}
m=\int^R_0\pi R l\left(4(\mu+\epsilon)+\frac{U}{\hat{E}}(q+\epsilon)B\right)dR+\frac{l^{2}\hat{Q}^2}{2R}-\frac{l^{2}}{2}\int^R_0\frac{\hat{Q}^{2}}{R^{2}}dR.
\end{equation}
Here, $m(0)=0$ has been used.
Now we obtain acceleration $D_{t}U$ as
\begin{equation}\label{9u}
D_{t}U=\frac{1}{A}\frac{\partial}{\partial t}\left(\frac{\dot{R}}{A}\right)
\Rightarrow D_{t}U=\frac{\ddot{R}}{A^{2}}-\frac{\dot{R}\dot{A}}{A^{3}}.
\end{equation}
Now from Eqs.(\ref{9f}), (\ref{9m}) and (\ref{9u}), we get
\begin{eqnarray}\label{9uu}
D_{t}U&=&-\left[\frac{m}{R^{2}}+8\pi\left(p+\Pi+\epsilon+\frac{2}{3}\Omega\right)R\right]+ \frac{A'\hat{E}}{AB}\\\nonumber
&+&\frac{\hat{Q^{2}}}{R}\left(\frac{l^{2}}{2R^{2}}-1\right)+\frac{l}{8R^{2}}\left(1+U^{2}-\hat{E^{2}}\right).
\end{eqnarray}
Using the conservation law, we get the following dynamical equations
\begin{eqnarray}\nonumber
T^{\mu\nu}_{;\nu}V_{\mu}&=&-\frac{1}{A}\left(\dot{\mu}+\dot{\epsilon}-\pi\dot{E}E\right)-\frac{1}{B}(\acute{q}+\acute{\epsilon})
-\frac{\dot{R}}{AR}\left(\mu+p+\Pi+\epsilon-\frac{\Omega}{3}-\frac{\pi}{2}E^{2}\right)\\\label{23} &-&\frac{\dot{B}}{AB}\left(\mu+p+\Pi+2\epsilon+\frac{2}{3}\Omega\right)-2\frac{(ABR)}{AB^{2}R}'(q+\epsilon)
\end{eqnarray}
and
\begin{eqnarray}\nonumber
T^{\mu\nu}_{;\nu}\chi_{\mu}&=&\frac{1}{A}\left(\dot{q}+\dot{\epsilon}\right)+\frac{2}{A}\frac{(BR)\dot{}}{B R}(q+\epsilon)
+\frac{1}{B}\left(\acute{p}+\acute{\Pi}+\acute{\epsilon}-\frac{\acute{2\Omega}}{3}-2\pi E\acute{E}\right)\\\label{24c} &+&\frac{1}{B}\frac{\acute{A}}{A}\left(\mu+p+\Pi+2 \epsilon+\frac{2}{3}\Omega\right)+\frac{2}{B}\frac{\acute{R}}{R}(\epsilon+\Omega).
\end{eqnarray}
Using the value of $\frac{A'}{A}$ from Eq.(\ref{9uu}) into Eq.(\ref{24c}) and considering Eqs.(\ref{1f})-(\ref{4f}), we get the main dynamical equation
\begin{eqnarray}\nonumber
\left(\mu+p+\Pi+2\epsilon+\frac{2}{3}\Omega\right)D_{t}U&=&-(\mu+p+\Pi+2\epsilon+\frac{2}{3}\Omega)\Bigg[\frac{m}{R^{2}}+8\pi R(p+\Pi+\epsilon+\frac{2}{3}\Omega)\\\nonumber
&-&\frac{\hat{Q}^{2}}{R}(\frac{l^{2}}{2R^{2}}-1)-\frac{l}{8R^{2}}(1+U^{2})\Bigg]\\\nonumber
&-&\hat{E^{2}}\left[D_{R}(p+\Pi+2\epsilon+\frac{2}{3}\Omega+\frac{\hat{Q}\acute{\hat{Q}}}{4\pi^{2}R})+\frac{\hat{Q}^{2}}{4\pi R^{3}}+\frac{2}{R}
(\epsilon+\Omega)\right]\\\label{s2}&-&
\hat{E}\left[D_{t}q+D_{t}\epsilon+2(q+\epsilon)\frac{U}{R}+\frac{2\dot{B}}{A B}(q+\epsilon)\right].
\end{eqnarray}
Here, we can analyze that the factor $\left(\mu+p+\Pi+2\epsilon+\frac{2}{3}\Omega\right)$ is common on the left side and first term on right side, this is the effective inertial mass, and according to equivalence principle it is also known as passive gravitational mass. On the right side, in the first term the square bracket factor explains the effects of dissipative variables on the active gravitational mass of the collapsing cylinder, this fact has been notified firstly by Herrera et al.\cite{39}. In the second square bracket there are gradient of total effective pressure which is influenced by dissipative variables, radiation density and electromagnetic field. The last square bracket contains different contributions due to dissipation nature of the system. The third term in this factor is positive $(U<0)$ implying the outflow of $q>0$ and $\epsilon>0$ reduces integrated energy of the contracting source, which decreases the rate of collapse.
\section{\textbf{The Transport Equation}}
The objective of this manuscript is to discuss a full causal approach for viscous dissipative gravitational collapse of stellar objects along with heat conduction. This implies that all dissipative variables must satisfy the transport equations obtained from causal thermodynamics. Consequently, we use the transport equations for heat, bulk and shear viscosity from the Muller-Israel-Stewart \cite{44}-\cite{46} for dissipative material. These transport equations for heat, bulk and shear viscosity \cite{39} are
\begin{eqnarray}\label{27a}
\tau_{0}\Pi_{;\alpha}V^{\alpha}+\Pi=-\xi\Theta+\alpha_{0}\xi q^{\alpha}_{;\alpha}-\frac{1}{2}\xi T \left(\frac{\tau_{0}}{\xi T}V^{\alpha}\right)_{;\alpha}\Pi,
\end{eqnarray}
\begin{eqnarray}\label{27}
\tau_{1}h^{\beta}_{\alpha}q_{\beta;\mu}V^{\mu}+q_{\alpha}&=&-\kappa \Bigg[h^{\beta}_{\alpha}T_{,\beta}(1+\alpha_{0}\Pi)+\alpha_{1}\pi^{\mu}_{\alpha}\\\nonumber &&h^{\beta}_{\mu}T_{,\beta}
+T (\alpha_{0}-\alpha_{0}\Pi_{;\alpha}-\alpha_{1}\pi^{\mu}_{\alpha;\mu}) \Bigg]-\frac{1}{}2
\kappa T^2\Bigg(\frac{\tau}{\kappa T^2}V^{\beta}\Bigg)_{;\beta}q_{\alpha},
\end{eqnarray}
\begin{eqnarray}\label{27b}
\tau_2 {h^{\mu}}_{\alpha} {\pi_{\mu\nu}}_{;\rho}V^{\rho}+\pi_{\alpha\beta}=-2\eta\sigma_{\alpha\beta}+2\eta\alpha_{1}q_{<\beta;\alpha>}-\eta T\Bigg(\frac{\tau^2}{2\eta T}V^{\nu}\Bigg)_{;\nu}\pi_{\alpha \beta},
\end{eqnarray}
\begin{eqnarray}\label{27c}
q_{<\beta;\alpha>}={h^{\mu}}_{\alpha}{h^{\nu}}_{\beta}\Bigg(\frac{1}{2}({q_{\mu}}_{;\nu}+{q_{\nu}}_{;mu})-\frac{1}{3}{q_{\sigma}}_{;\kappa}h^{\sigma\kappa}\Bigg),
\end{eqnarray}
where relaxation times have following values
\begin{equation}
\tau_0=\xi \beta_0\quad\quad \tau_1=\kappa T \beta_1\quad\quad \tau_2=2\eta\beta_2,
\end{equation}
where $\beta_1$,\quad $\beta_2$ are thermodynamic coefficients for different contributions to entropy density, $\alpha_0$,\quad $\alpha_1$ are thermodynamics viscous/heat coupling coefficients, $\xi$ and $\eta$ are coefficients of bulk and shear viscosity. The Eq.(\ref{27a})-(\ref{27c}), with the help of given interior metric take the following form

\begin{eqnarray}\label{27}
\tau_{0}\dot{\Pi}=-(\xi+\frac{\tau_{0}\Pi}{2})A\Theta+\frac{A}{B}\alpha_{0}\xi\left[\acute{q}+q(\frac{\acute{A}}{A}+\frac{2\acute{R}}{R})\right]
-\Pi\left[\frac{\xi T}{2}(\frac{\tau_{0}}{\xi T})\dot{}+A\right],
\end{eqnarray}
\begin{eqnarray}\nonumber
\tau_{1}\dot{q}&=&\frac{A}{B}\kappa\acute{T}(1+\alpha_{0}\Pi+\frac{2}{3}\alpha_{1}\Omega)+T[\frac{\acute{A}}{A}-\alpha_{0}\acute{\Pi}-\frac{2}{3}
\alpha_{1}(\acute{\Omega}+(\frac{\acute{A}}{A}+3\frac{\acute{R}}{R})\Omega)]\\\label{28aa}
&-&q\left[\frac{\kappa T^2}{2}(\frac{\tau_{0}}{\kappa T^2})\dot{}+\frac{\tau_{1}}{2}\Theta A+A\right],
\end{eqnarray}
\begin{eqnarray}\label{29}
\tau_{1}\dot{\Omega}&=&-2\eta\sigma+2\eta\alpha_{1}\frac{A}{B}(\acute{q}-q\frac{\acute{R}}{R})-\Omega\left[\eta T\left(\frac{\tau_{2}}{2\eta T}\right)\dot{}+\frac{\tau_{2}}{2}\Theta A+A\right].
\end{eqnarray}
Now, to observe the influence of various dissipative variables on cylindrical collapsing source, we substitute Eq.(\ref{28aa}) in Eq.(\ref{s2}) and after some rearrangements, we obtain

\begin{eqnarray}\nonumber
(\mu+p+2\epsilon +\frac{2}{3}\Omega)(1-\Lambda)D_{t}U&=&(1-\Lambda)F_{grav}+F_{hyd}\\\nonumber
&+&\frac{\kappa}{\tau_{1}}{\hat{E}}^{2}\left[D_{R}T(1+\alpha_{0}\Pi+\frac{2}{3}\alpha_{1}\Omega)\right]\\\nonumber
&-&\frac{\kappa}{\tau_{1}}{\hat{E}}^{2}T\left[\left(\alpha_{0}D_{R}\Pi+\frac{2}{3}\alpha_{1}+
(D_{R}\Omega+\frac{3}{R}\Omega)\right)\right]\\\nonumber
&-&\hat{E}\left[\frac{2\dot{B}}{AB}(q+\epsilon)-\frac{q}{\tau_{1}}-2(q+\epsilon)\frac{U}{R}\right]\\\label{27aaa} &+&\hat{E}\left[\frac{\kappa, T}{2\tau_{1}D_{t}}(\frac{\tau_{1}}{\kappa T^{2}}-D_{t}\epsilon)+A\frac{\tau_{1}}{2}\Theta\right]
\end{eqnarray}
where $F_{grav}$, $F_{hyd}$ and $\Lambda$ defined by

\begin{eqnarray}\nonumber
F_{grav}&=&-(\mu+p+\Pi+2\epsilon+\frac{2}{3}\Omega)\Bigg[\frac{m}{R^{2}}+8\pi R(p+\Pi +\epsilon+\frac{2}{3}\Omega)-\frac{\hat{Q}^{2}}{R}(\frac{l^{2}}{2R^{2}}-1)\\\nonumber &-&\frac{l}{8R^{2}}(1+U^{2})\Bigg]\\\nonumber
F_{hyd}&=&-\hat{E^{2}}\Bigg[D_{R}\Bigg(p+\Pi+\epsilon+\frac{2}{3}\Omega+\frac{\hat{Q}^{2}}{4\pi^{2}R^{2}}-\frac{\hat{Q}}{2\pi^{2}R^{2}}\Bigg)
+\frac{\hat{Q}^{2}}{2\pi^{2}R^{3}}+\frac{2}{3R}(\epsilon+\Omega)\Bigg]\\\label{28}
\Lambda&=&\frac{\kappa T}{\tau_{1}}\Bigg(\mu+p+2\epsilon+\frac{2}{3}\Omega\Bigg)^{-1}\Bigg(1-\frac{2}{3}\alpha_{1}\Omega\Bigg).
\end{eqnarray}
Taking value of $\Theta$ from Eq.(\ref{27a}) and using Eq.(\ref{27aaa}), we have the following resulting equation
\begin{eqnarray}\nonumber
(\mu+p+2\epsilon +\frac{2}{3}\Omega)(1-\Lambda+\Delta)D_{t}U&=&(1-\Lambda+\Delta)F_{grav}+F_{hyd}\\\nonumber
&+&\frac{\kappa}{\tau_{1}}{\hat{E}}^{2}\left[D_{R}T(1+\alpha_{0}\Pi+\frac{2}{3}\alpha_{1}\Omega)\right]\\\nonumber
&-&\frac{\kappa}{\tau_{1}}{\hat{E}}^{2}T\left[\left(\alpha_{0}D_{R}\Pi+\frac{2}{3}\alpha_{1}
(D_{R}\Omega+\frac{3}{R}\Omega)\right)\right]\\\nonumber &-&\hat{E}^{2}\left(\mu+p+2\epsilon+\frac{2}{3}\Omega\right)\Delta\left(\frac{D_{R}q}{q}-\frac{4q}{R}\right)\\\nonumber
&-&\hat{E}\left[\frac{2\dot{B}}{AB}(q+\epsilon)-\frac{q}{\tau_{1}}-2(q+\epsilon)\frac{U}{R}\right]\\\nonumber &+&\hat{E}\left[\frac{\kappa T^2q}{2\tau_{1}}D_{t}(\frac{\tau_{1}}{\kappa T^{2}})-D_{t}\epsilon\right]\\\nonumber&+&\hat{E}\left(\mu+p+2\epsilon+\frac{2}{3}\Omega\right)\frac{\Delta}{2\alpha_0\kappa q}\\\label{last}
&\times&\left(1+2\xi T D_{t}(\frac{\tau_{0}}{\xi T})\Pi+\frac{\tau_{0}}{A}D_{t}\Pi\right),
\end{eqnarray}

where
\begin{eqnarray}
\Delta=\alpha_0\xi q\Bigg(\frac{3q+4\epsilon}{2\xi+\tau_0\Pi}\Bigg)\left(\mu+p+\Pi+2\epsilon+\frac{2}{3}\Omega\right)^{-1}.
\end{eqnarray}
Hence, by taking into account the casual transportation equations and their coupling with the dynamical equations, we find that factor $(\Delta-\Lambda+1)$ affects significantly the internal energy and passive gravitational mass density. This result is in the agreement with \cite{39}. We would to mentioned that we have considered the charged dissipative viscous fluid, but there appears no terms in the relevant equations which couples the electromagnetic field and dissipative variables. Therefore the role of electromagnetic field on the dynamical process is same as in the absence of shear viscosity already discussed in \cite{sm1}.

\section{Conclusion}
%
 Immediately after the Einstein's theory of gravity in early of 20th century, the study of of cylindrically symmetric systems was started by Weyl \cite{W1} and Levi-Civita \cite{L1}. After the complete study of spherical objects, theoretical physicists were interested to explore the properties of astrophysical compact stars that have axially symmetry. The relativistic fluids involving heat flux and viscosity are very important for studying the evolution of compact. So, it is important to include the dissipative variables in gravitational collapse.

In this paper, we have constructed the dynamical equations which deal significantly with the structure and evolutionary phases of a charged gravitating viscous cylindrical source. In order to see the effects of dissipation on the dynamical evolution of gravitating source, we have assumed the convenient form of the dissipative variables which satisfy the transportation equations of heat, bulk and shear viscosity resulting from the casual thermodynamics. Furthermore, the dissipative coefficients due to viscosity and heat flux have been included in the discussion of dynamical equations. In a very broad sense, we are mainly interested in time scales whose order may be smaller or equal to the radiation time. During the study of transport equations for dissipative variables, we have preferred to use the hyperbolic theory of dissipation because this theory is more reliable than that of parabolic theory and have less difficulties as in parabolic theory arise \cite{8},\cite{46}-\cite{48}.

A full casual approach has been adopted in \cite{39} to explore the effects of dissipative variables on the spherical spherical collapse, this study provide the meaning full results which as significant implications in astronomy. The application of  these results to some stellar system implies that in a pre-supernovae event, thermal conductivity of the dissipative source might be large enough to produce an observable reduction in gravitational force of the system that results to the expansion of gravitating source instead of collapse. It is quite relevant to mentioned that thermodynamics viscous/heat coupling coefficients have been taken as non-vanishing because this assumption provide the significant basis for the modeling of non-uniform stellar system. In a recent investigation \cite{sm1}, we have considered non-casual (irreversible thermodynamics) approach to discuss the dynamics of charged bulk viscous cylindrical collapse by neglecting the thermodynamics viscous/heat coupling coefficients in the transportation equations. So, our present analysis is the extension of our previous study with non-casual approach \cite{sm1}. \textit{But it is important to note that this analysis with cylindrical symmetry is analogous to full casual approach adopted by Herrera et al.\cite{39} for spherical stellar objects}.

As a consequence of a full casual approach to the dynamics of charged dissipative cylindrical collapse, we obtain a dynamical equation (\ref{last}), which explains how the value of effective inertial mass is influenced by the dissipative variables and thermodynamics viscous/heat coupling coefficients. All the dissipative variables have a great effect in the a pre-supernovae event, for example the large enough value of heat conductivity $\kappa$, can produce a rapid decrease in the force of gravity thereby resulting in the reversal of collapse \cite{39}. A numerical model predicting this type of bouncing behavior has been presented Herrera et al.\cite{11}. It is to be noted that such numerical estimations in the present case are beyond the scope of this this work. Here we just want to ensure that during the dissipative gravitational collapse, thermodynamics viscous/heat coupling coefficients must not be excluded priori in the transportation equations. In future, we are interested to extend these results in modified $f(R)$, $f(R,T)$ and $f(G)$ theories of gravity.

\end{document}